# Lunar impact craters identification and age estimation with Chang'E data by deep and transfer learning

Chen Yang[1,2*], Haishi Zhao[3], Lorenzo Bruzzone[4], Jon Atli Benediktsson[5], Yanchun Liang[3], Bin Liu[2], Xingguo Zeng[2], Renchu Guan[3*], Chunlai Li[2*] & Ziyuan Ouyang[2,6]

**Impact craters, as "lunar fossils", are the most dominant lunar surface features and occupy most of the Moon's surface. Their formation and evolution record the history of the Solar System. Sixty years of triumphs in the lunar exploration projects accumulated a large amount of lunar data. Currently, there are 9137 existing recognized craters. However, only 1675 of them have been determined age, which is obviously not satisfactory to reveal the evolution of the Moon. Identifying craters is a challenging task due to their enormous difference in size, large variations in shape and vast presence. Furthermore, estimating the age of craters is extraordinarily difficult due to their complex and different morphologies. Here, in order to effectively identify craters and estimate their age, we convert the crater identification problem into a target detection task and crater age estimation into a taxonomy structure. From an initial small number of available craters, we progressively identify craters and estimate their age from Chang'E data by transfer learning (TL) using deep neural networks. For comprehensive identification of multi-scale craters, a two-stage craters detection approach is developed. Thus 117,240 unrecognized lunar craters that range in diameter from 532 km to 1 km are identified. Then, a two-stage classification approach is developed to estimate the age of craters by simultaneously extracting their morphological features and stratigraphic information. The age of 79,243 craters larger than 3 km in diameter is estimated. These identified and aged craters throughout the mid and low-latitude regions of the Moon are crucial for reconstructing the dynamic evolution process of the Solar System.**

The Moon's surface is full of impact craters for lack of atmosphere and water with little erosion and geological effects. The age of craters on the Moon spans five age Systems, i.e. the pre-Nectarian System (3.92-4.52 Ga), the Nectarian System (3.85-3.92Ga), the Imbrian System (3.16-3.85Ga), the Eratosthenian System (0.8-3.16Ga) and the Copernican (0-0.8 Ga) System, and is as long as four billion years. Their formation and evolution record the history of the Solar System [1-5]. As early as 400 years ago, people first became aware of lunar impact craters. Sixty years of triumphs in the lunar exploration projects (e.g., the Luna missions and NASA's Apollo programme) accumulated a large amount of lunar data, including digital images, digital elevation models (DEM), a database of lunar craters and lunar geologic maps. According to statistics, the total number of craters larger than 10km in diameter on the moon over 33,000[6]. The international recognized and named craters[7] regulated by the International Astronomical Union (IAU) since 1919 are 9137. The age of only 1675 of these craters is constrained age and aggregated by the Lunar and Planetary Institute (LPI)[8], mainly according to a professional paper from the United States Geological Survey (USGS). That paper is based on the geologic history of the Moon[9], the stratigraphy of lunar craters database[10] and geologic map of the near and east side of the Moon[11,12]. Fig. 1a shows the distribution of recognized craters on the Moon. Obviously, existing recognized and age constrained craters are not adequate enough to reveal their evolutionary history and process.

Lunar craters have the same genesis, i.e., impacts create craters that look similar in a near-circular depression structure. This is the main basis for the identification of craters. Different experiences, i.e., the formation and long-term alteration, lead to craters having a different complex morphology. Typical characteristics can demonstrate differences in orders of magnitude in size of the diameters, e.g. the largest craters have a diameter of a few hundred kilometers, whereas the smallest ones have a diameter of a few meters. They also show large variations in shape due to an overlap with other craters (see Fig.1a). Traditional manual identification becomes a tedious process, which can be considered only practical for a few comprehensive craters in specific geographic regions. Existing automatic detection algorithms[13-15] that are based on pattern recognition and machine learning (ML) can determine the vast presence of small simple craters from the general features of craters. The deep learning paradigm and in particular deep convolutional neural networks (CNN) can perform fine-grained information extraction, demonstrating fast and accurate performances[16,17]. Nevertheless, deep learning requires very large datasets[18,19] with huge amounts of detected craters as labeled samples for training. Such large quantities of samples are not available. Moreover, the available samples mainly refer to simple craters and thus cannot model seriously degraded craters with irregular shapes that may have formed in early periods and provide an important history record. From another perspective, craters formed in a given geological period display different characteristics for different ages (see Fig.1a). However, existing craters that have received an age estimation by astronomers based on morphological markers[20,21] and stratigraphic analysis[22], are only one fifth of the recognized craters. This is due to the fact that the slow and long-term evolution makes the morphologies of craters of different ages complex and confusing resulting in an extraordinarily difficult age estimation task.

Facing the distinctively complex nature of craters, it is difficult to obtain satisfactory results in crater identification and age estimation with a single type of data using conventional techniques. In the new generation of exploration on the Moon, the Chang'E-1 and Chang'E-2 orbiters[23] of the China's Lunar Exploration Program (CLEP) have enriched lunar data, acquiring two different spatial resolution images, i.e., 120m and 7m digital orthophoto images (DOM) (the highest resolution of a global digital image map of the Moon so far), and 500m and 7m DEM data covering the whole lunar surface. Craters exhibit different features under different resolutions. Large-scale data with a low-resolution present the morphology of large craters, while the high resolution data are important for small craters. Despite the fact that these data contain the required information for crater analysis, a large number of craters have not been detected and/or obtained age estimation. Transfer learning (TL)[24], one of the frontiers of ML,

[1]College of Earth Sciences, Jilin University, Changchun, China. [2]Key Laboratory of Lunar and Deep Space Exploration, National Astronomical Observatories, Chinese Academy of Sciences, Beijing, China. [3]College of Computer Science and Technology, Jilin University, Changchun, China. [4]Department of Information Engineering and Computer Science, University of Trento, Trento, Italy. [5]Electrical and Computer Engineering, University of Iceland, 101 Reykjavik, Iceland. [6]Institute of Geochemistry, Chinese Academy of Sciences. *e-mail: yangc616@jlu.edu.cn; guanrenchu@jlu.edu.cn; licl@nao.cas.cn  *These authors contributed equally to this work.

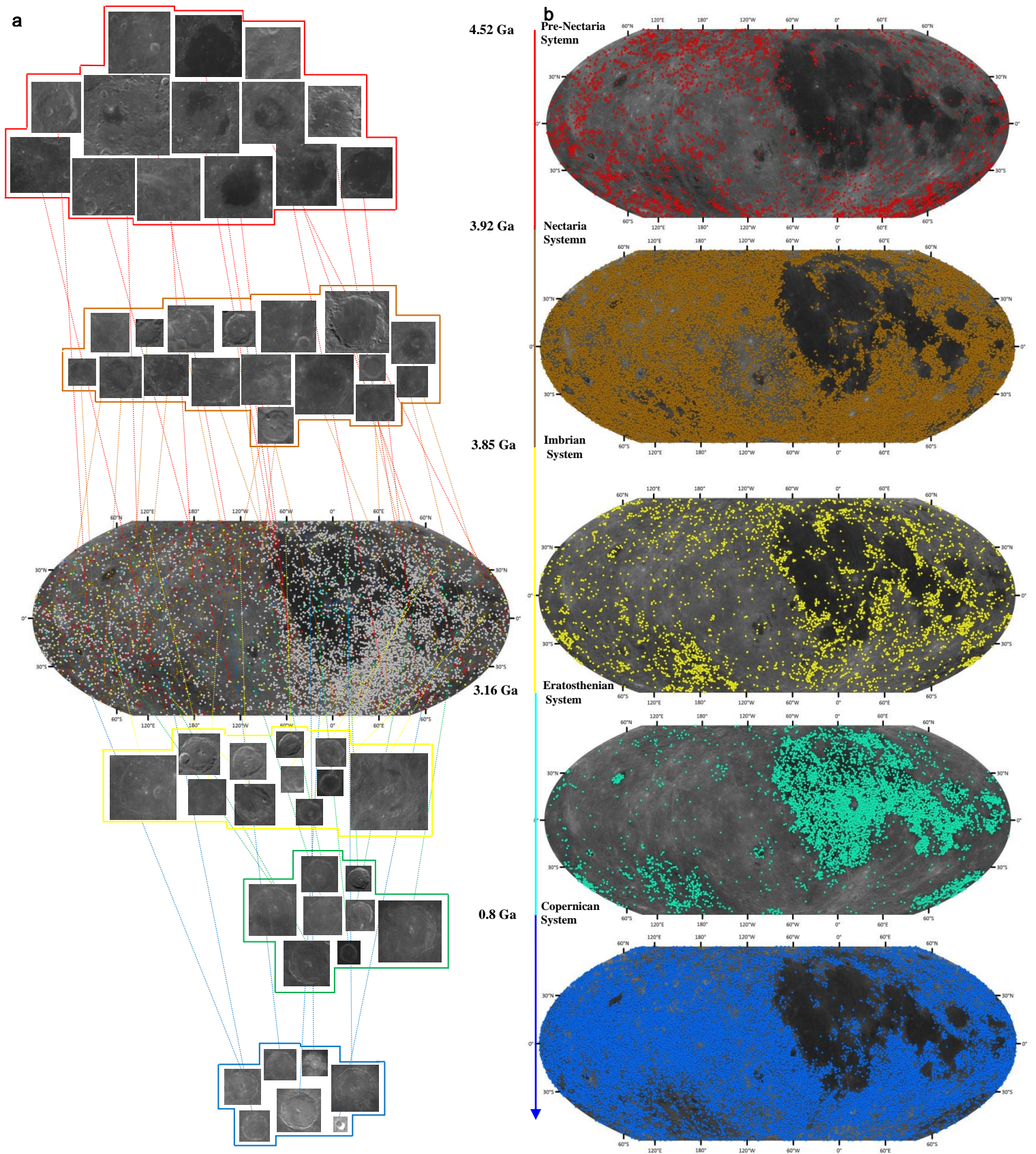

**Fig. 1 | Distribution of lunar impact craters on the Moon. a,** The distribution of recognized and aged craters. The red, brown, yellow, green and blue squares and points represent the craters of the pre-Nectarian System, the Nectarian System, the Imbrian System, the Eratosthenian System and the Copernican System, respectively. The gray points show the recognized craters without ages. **b,** Distribution of identified and assigned age craters. From time-scale and the space distribution, these aged craters exhibit specific characteristics. The craters having diameter smaller than 3 km and larger than 600km are not shown in the distribution map.

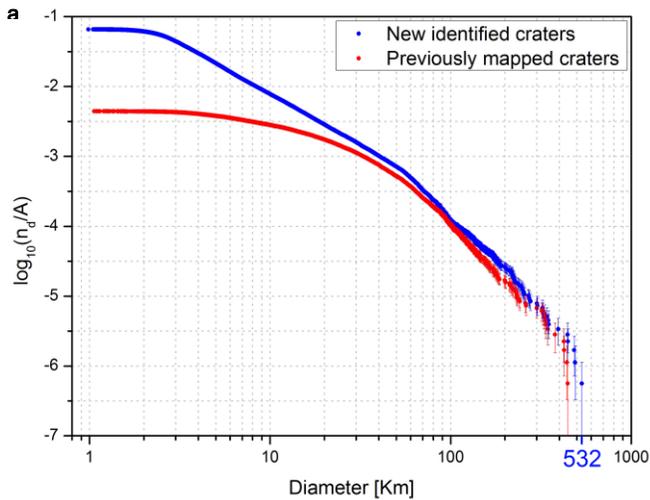 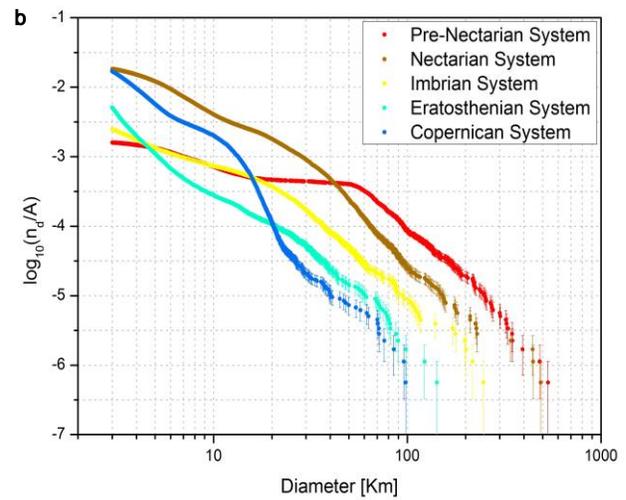

**Fig. 2 | Lunar impact crater size-frequency distributions with CSFD plots.**
**a,** The CSFD of craters with the new identified and the known craters. The blue line represents the identified craters compared with the known craters (red line). The craters having diameters smaller than 1 km and larger than 500km are not used for identification due to resolution.

**b,** The CSFD of lunar impact craters with different Systems. The red, brown, yellow, green and blue lines show craters CSFD of the pre-Nectarian System, the Nectarian System, the Imbrian System, the Eratosthenian System and the Copernican System, respectively.

motivated by the fact that humans apply previous knowledge to solve new problems intelligently[25], is successfully applied to the problems where sufficient training samples are not available [26,27]. In this paper, we progressively identify craters and estimate their ages with Chang'E-1 and Chang'E-2 data by means of TL. 117,240 unidentified craters that range from 532 km to 1 km in diameter are detected, and the age of 79,243 craters larger than 3 km is estimated. The distribution map of identified and aged lunar craters is presented in Fig.1b. The craters span the full range in latitude -180° to 65°, -65° to 65° and 65° to 180° in longitude on the Moon (the mid and low-latitude regions).

The comprehensive identification of multi-scale lunar impact craters is based on the proposed two-stage crater detection approach with Chang'E-1 and Chang'E-2 data..

Considering the magnitude difference of the craters scale, a series of lunar craters images are taken with Chang'E-1 and Chang'E-2 data corresponding to different spatial resolutions and sizes from different angles. These craters images are obtained by the fusion of digital orthophoto map (DOM) data and DEM data (see Methods). The former presents the morphological characteristics, whereas the latter displays the topographic information of craters. Three scales of images, i.e., 120m Chang'E-1 images with 5000×5000 and 1000×1000 pixels, as well as 50m Chang'E-2 images with 1000×1000 pixels, are formed for detecting craters having large, medium, and small diameter ranges, i.e., 600km-36km, 120km-20km, and 50km-0.1km. Adjacent images have a 50% overlap, and each crater may appear in two or three images.

In the first stage of detection approach, recognized craters in Chang'E-1 images are randomly divided into three separate datasets, i.e., 5682, 1422 and 791 images for training, validation and test sets, and all recognized craters in Chang'E-2 images (i.e., 6511 images) are used for testing the second stage of the detection model. The first stage of detection approach achieves 94.71% recall, recovering almost all the recognized craters in the test set and detecting a large number of new craters. Then the first stage detection model is directly transferred to Chang'E-2 images obtaining a 93.35% recall. The average detection time required for each image is 0.17 seconds (see Methods).

Fig. 2a shows the cumulative size-frequency distribution (CSFD) of craters identified by the detection model compared with the previously mapped craters. It can be observed that the CSFD of the identified craters is systematically higher than that of the known craters for diameters between 1km and 100 km. This indicates that the detection approach finds a substantially larger number of craters than those previously mapped in the small and medium diameter ranges. From 100 km to 532km, the curves of the identified and the previously mapped craters have a very similar behavior. This is due to the fact that the large craters are sparse on the Moon.

The specific System defined for the Moon was based on the major impact events and erosion morphologies of craters. For example, the pre-Nectarian System includes all deposits that are older than Nectaris Basin; the Nectarian System and the Imbrian System started with the Nectaris and Imbrium impact events, respectively. The Eratosthenian System is the period in which craters can be clearly distinguished but the bright sputtering materials, i.e., crater rays around those craters, begin to darken and disappear. The Copernican System is defined by craters with bright rays which present recent lunar geologic record. Therefore, craters ages are estimated by observation with stratigraphic technique, e.g., the position of ejecta blankets and their morphologies, (i.e., the degree of erosion).

In this paper, the five Systems, i.e. the pre-Nectarian System, the Nectarian System, the Imbrian System, the Eratosthenian System and the Copernican System are mapped into a taxonomy structure. Then, as in the previous case, a two-stage crater classification approach based on TL with Chang'E-1 and Chang'E-2 data is proposed. To take advantage of morphology features and stratigraphic information of the scarce aged and enormous identified craters, a semi-supervised dual-channel craters classification model is established (see Methods).

In the first stage of classification approach, 1411 craters with constrained ages in Chang' E-1 were associated with the training, the validation and the test sets with the proportions of 8:1:1. 502 craters in Chang'E-2 images are used for testing the second stage classification. The effectiveness of the classification approach in the first stage is validated with Chang'E-1 data on five trials. In the first stage of classification, the classification approach obtains 84.12 ± 2.68% (mean ± s.d.) overall accuracy and achieves the best overall accuracy of 86.05% on a subset of aged craters with diameters between 500km and 3km. This demonstrates that this classification model trained in the first stage of classification approach has the ability to accurately classify craters into their respective Systems. The best performance model in the first stage is transferred to Chang'E-2 data without training, resulting in 88.05% of aged craters with diameters between 50km and 1km being classified correctly. Finally, the classification approach is utilized to assign ages to identified craters and previously mapped craters without ages. The average time of the classification of each crater is 0.006 second (see Methods). The spatial distribution of all aged craters (larger than 3 km in diameter for avoiding the effects of secondary craters) on the Moon is shown in Fig. 1b.

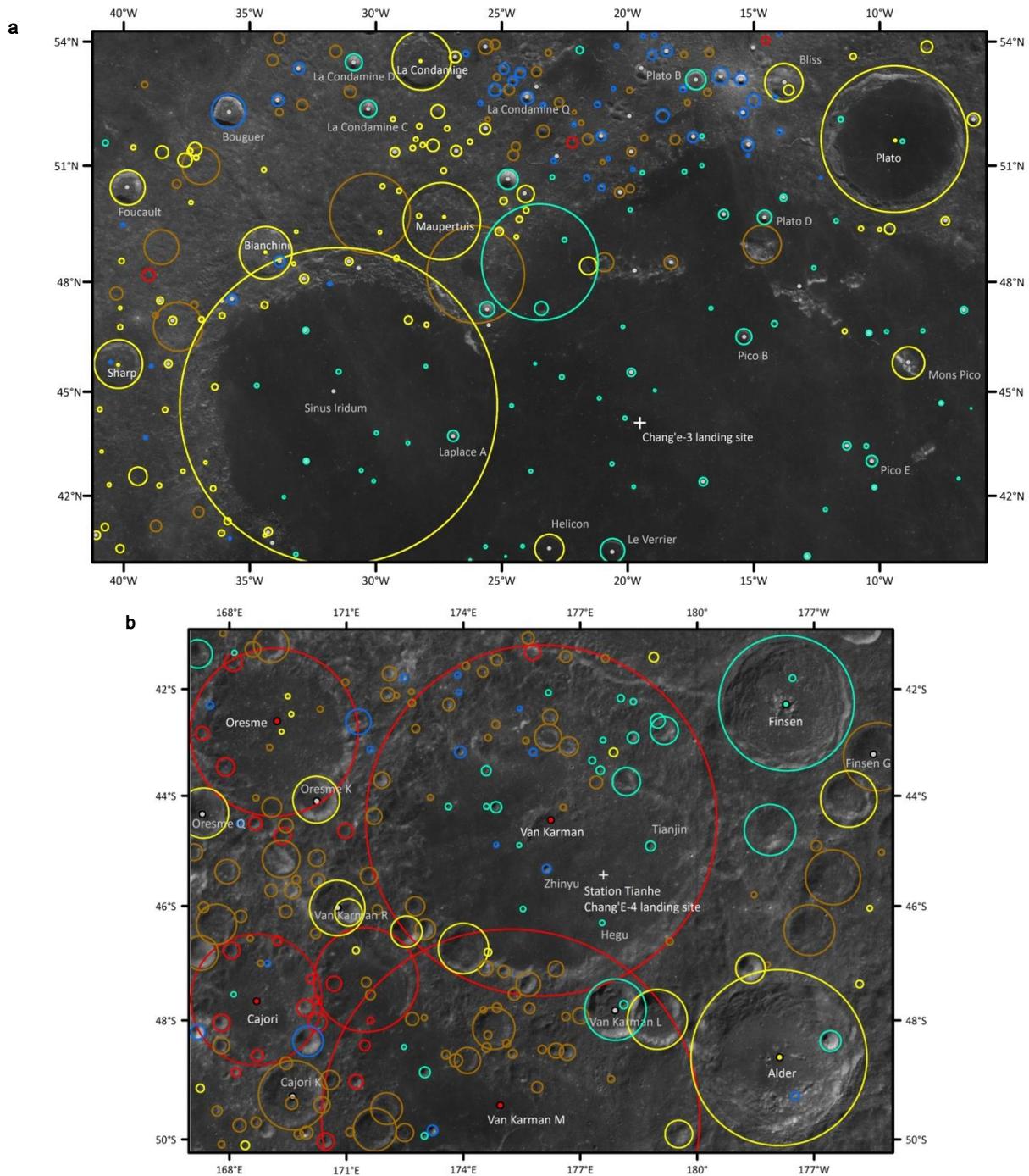

**Fig. 3 | Aged lunar impact craters at typical areas on the front and back of the Moon. a,** Craters with constrained ages in north of the Mare Imbrium. This area contains 76 previous mapped craters (5 with constrained ages) and 262 identified craters. **b,** Aged craters in northwest of the South Pole-Aitken (SPA) basin. There are 12 previous mapped craters (6 with constrained ages) and 197 identified craters. The presented craters are larger than 3 km in diameter. The names of the previous mapped and aged craters are marked in white and those without ages are marked in gray. The red, brown, yellow, green and blue lines show aged craters of the pre-Nectarian System, the Nectarian System, the Imbrian System, the Eratosthenian System and the Copernican System, respectively.

The CSFDs of craters with five Systems are given in Fig. 2b. It can be seen that there are significant differences between the distributions of craters in diameters in the five Systems. For the pre-Nectarian System, the CSFD curve at first decreases slowly by increasing the diameter; then it begins to sharply decline at 70km, and the diameters range between 3km and 532km. The CSFD curve of the Nectarian System is sharply higher than that of the pre-Nectarian System when the diameters range from 3km to 60km, whereas the number of craters increases slowly in the diameters range between 60 km and 300km, and the largest crater has a diameter of roughly 500km. The Imbrian System curve decreases gradually but cuts off sharply at diameters around 200 km. The curve of the Eratosthenian System declines sharply at 100km in diameter. The Copernican System curve represents a fluctuating downward trend, and does not show diameters larger than 100 km. For diameter ranges between 3km and 10 km, the Nectarian System and the Copernican System CSFD curves are higher than those of the other Systems. Between 10km and 100km, the pre-Nectarian System and the Nectarian System show higher CSFD values. Between 100km and 532km, the pre-Nectarian System has the highest CSFD curve. Finally, the CSFD curves of the five Systems run essentially parallel to each other for diameters larger than 70km. These identification and classification results reveal the evolution of the Moon. The spatial distribution and the number of craters vary greatly in the five different Systems. This provides insights into the dynamical history of the Solar System.

The pre-Nectarian System (3.92-4.52 Ga), the oldest lunar geologic time, covers the period from the Moon's origin and formation of the lunar crust to the Nectaris impact. Craters formed in this period can provide on the evolution of the early Solar System. According to the record, the Moon was heavily impacted by asteroids and other celestial bodies in the Solar System during this period, and craters morphologies are characterized by long-term erosion. Results show that the craters of the pre-Nectarian System are large in diameter (the number of craters larger than 50km is far more than those of other Systems). There are 2847 craters in the study areas, about 715 larger than 50km. As one can see in Fig.1b, craters of the pre-Nectarian System are widely distributed in the south and north and in a larger area on the back of the Moon in the mid and low-latitude regions.

The Nectarian System (3.85-3.92Ga) is still in a very high impact environment. A famous hypothesized event, i.e., the Late Heavy Bombardment (LHB) also called the "lunar cataclysm", occurred approximately 4 billion years ago. During this period a large number of impact craters were formed on the Moon. Data from lunar samples (Apollo, Luna, and lunar meteorites) indicate that the Moon was subjected to an intense period of bombardment around 3.85 billion year ago (Ga)[28]. However, only 646 of the previous mapped craters were associated to the Nectarian System. The number of identified craters in the study area is the highest, i.e., 32,714. It is consistent with and further proves the LHB occurred on the Moon. The identified craters are mainly located on the back of the Moon. Craters of the Nectarian System with diameter smaller than 20km are more than those in the pre-Nectarian System.

In the Imbrian System (3.16-3.85Ga), many large-scale basaltic eruptions occurred after the formation of lunar mare, which refer to the "lunar mare flooding". The majority of mare basalts appear to have erupted between about 3 and 3.5 Ga[29]. In our study area, the number of craters is 4415, which is one seventh of the impact craters in the Nectarian System. They are mainly distributed in the front of the Moon and around lunar mare, most of the identified craters are distributed above the ejecta of the mare and filled with mare basalts.

The Eratosthenian System (0.8-3.16Ga) is a relatively young geological period on the lunar surface. The craters formed during this period were basically preserved, most of the crater rays had been eroded and destroyed. In the study area, the number of detected craters is 9083. They are mainly distributed in the medium and high $TiO_2$[30] and $FeO$[31] basalt region of the Mare Imbrium. From the chemical compositions of rocks and surface ages of mare basaltic units in Mare Imbrium estimated with Chang'E-1 and Clementine UVVIS data, the evolution of the basalts is from low-titanium and low-iron to high-titanium and high-iron from the Imbrian System to the Eratosthenian System[32]. The impact in the Eratosthenian System maybe the external cause of lunar multistage volcanic eruption in the Imbrium basin.

The Copernican System (0-0.8 Ga) is the youngest lunar geologic time and was once considered a period of not much crater formation activity. Our results show that the craters formed in this period are scattered all over the lunar surface in the study area. The number of craters is 30,184. Some craters formed during this period can be used to trace the history of the Earth. Latest researches[33] pointed out that the lunar impact rate has increased by a factor of 2.6 in the past 290 Ma compared with the preceding ~710 Ma.

Two typical regions, i.e., the north of Mare Imbrium on the front of the Moon (the Chang'E-3[34,35] landing site) and the northwest of the South Pole-Aitken (SPA) basin located in the back of the Moon, are selected for further analysis (Fig. 3). The Mare Imbrium is the largest mare associated with the impact basin and is the beginning of the Imbrian System. From Fig. 3a, we can see that the Sinus Iridum, i.e., "Bay of Rainbows" is located in the northwest corner of Mare Imbrium and is aged at the Imbrian System by using the crater size-frequency distributions (CSFDs), which is consistent with the conclusion of some researchers[36]. The Imbrium System craters are distributed around the Mare Imbrium. Most craters in the Mare Imbrium are related to the Eratosthenian System. Fig. 3b shows the Von Kármán crater[37,38] (more than 180 kilometers in diameter), which is the oldest crater in the Solar System. Chang'E-4 landing site (Tianhe) is located in this crater. Three craters are arranged in triangle around the Tianhe, i.e., Tianjin and Hegu are aged with the Eratosthenian System, while Zhinyu is assigned to the Copernican System.

In this paper, we progressively identified the lunar impact craters and estimated their ages with Chang'E-1 and Chang'E-2 data based on transfer learning by means of deep neural networks. Through a two-stage detection and classification approaches, an enormous amount of craters were identified and their ages estimated starting from the limited number of already mapped craters. The results reveal the main distribution of existing craters (ranging from 532km to 1km in diameter) in the mid and low-latitude regions on the Moon after its origin. Craters having different ages show their specific distribution and characteristics on the front and back of the Moon, on lunar mare and highland in lunar mid and low-latitude regions. This provides significant evidence of the evolution of the Moon. Meanwhile, the periodic changes in the number of craters in the five ages reflect the dynamic history of Solar System. In summary, these large number of identified and assigned ages craters bring us new information and understanding about the evolution history of the Moon and the Solar System. Further research is to transfer the detection approach to the Chang' E-2 7m data for identifying even smaller sub-kilometer craters in order to build a lunar craters database and provide more elaborate information in the lunar exploration. This progressive transfer learning strategy implemented in a deep architecture is like a "Supervisor" passing his knowledge (feature representation) and experience (classification capability) from one generation to another. The "Student" learned model (e.g. the two-stage crater detection approach) could be adapted to other Solar System bodies, e.g. Mars, Mercury, Venus, Vesta and Ceres to extract much more semantic information with respect to the usually manually analyzed data.